\documentclass[aps,prd,nofootinbib,superscriptaddress,prd,aps,floatfix]{revtex4}
\usepackage[latin9]{inputenc}
\usepackage{color,textcomp,amsmath,amssymb,graphicx,amsfonts,graphics,epstopdf,slashed,multirow,adjustbox,lipsum,rotating,xcolor,wrapfig,epsfig}
\usepackage[unicode=true, bookmarks=false, breaklinks=false,pdfborder={0 0 1},backref=section,colorlinks=false]{hyperref}
\hypersetup{colorlinks,bookmarksopen,bookmarksnumbered,linkcolor=blu,pdfstartview=FitH,urlcolor=rossos,citecolor=rossos}
\usepackage[justification=raggedright, margin=5mm,font={sf,small},labelfont=bf, format=plain]{caption}

\def\simlt{\stackrel{<}{{}_\sim}}
\def\simgt{\stackrel{>}{{}_\sim}}

\definecolor{rosso}{cmyk}{0,1,1,0.4}
\definecolor{rossos}{cmyk}{0,1,1,0.55}
\definecolor{rossoc}{cmyk}{0,1,1,0.2}
\definecolor{blu}{cmyk}{1,1,0,0.3}
\definecolor{blus}{cmyk}{1,1,0,0.6}
\definecolor{bluc}{cmyk}{1,1,0,0.1}
\definecolor{verde}{cmyk}{0.92,0,0.59,0.25}
\definecolor{verdec}{cmyk}{0.92,0,0.59,0.15}
\definecolor{verdes}{cmyk}{0.92,0,0.59,0.4}

\begin{document}

\title{Charged Lepton Flavor Violation in a class of Radiative Neutrino Mass Generation Models}

\author{Talal Ahmed Chowdhury}
\email{talal@du.ac.bd}
\affiliation{Department of Physics, University of Dhaka, P.O. Box 1000,
Dhaka, Bangladesh.}
\affiliation{The Abdus Salam International Centre for Theoretical Physics, Strada
Costiera 11, I-34014, Trieste, Italy.}

\author{Salah Nasri}
\email{snasri@uaeu.ac.ae}
\affiliation{Department of Physics, UAE University, P.O. Box
17551, Al-Ain, United Arab Emirates}
\affiliation{The Abdus Salam International Centre for Theoretical Physics, Strada
Costiera 11, I-34014, Trieste, Italy.}

\begin{abstract}
We investigate charged lepton flavor violating processes $\mu\rightarrow e \gamma$, $\mu\rightarrow e e \overline{e}$ and $\mu-e$ conversion in nuclei for a class of three-loop radiative neutrino mass generation models with electroweak multiplets of increasing order. We find that, because of certain cancellations among various one-loop diagrams which give the dipole and non-dipole contributions in effective $\mu e \gamma$ vertex and Z-penguin contribution in effective $\mu e Z$ vertex,  the flavor violating processes $\mu\rightarrow e\gamma$ and $\mu-e$ conversion in nuclei become highly suppressed compared to $\mu\rightarrow e e \overline{e}$ process. Therefore, the observation of such pattern in LFV processes may reveal the radiative mechanism behind neutrino mass generation.
\end{abstract}

\pacs{04.50.Cd, 98.80.Cq, 11.30.Fs.}
\maketitle

\section{Introduction}\label{intro}

Although we have observed lepton flavor violation (LFV) in the neutral fermion sector of the Standard Model (SM) in neutrino oscillation, the charged LFV in the SM has turned out to be highly suppressed. For example, by allowing massive neutrinos, $m_{\nu}\sim 1$ eV and leptonic mixing matrix, Pontecorvo-Maki-Nakagawa-Sakata (PMNS) matrix, $U_{\text{PMNS}}$ in the SM, the branching ratio of charged lepton violating process, $\mu\rightarrow e\gamma$, turns out to be about $10^{-47}$ \cite{Cheng:1976uq, Cheng:1977nv, Petcov:1976ff, Marciano:1977wx, Lee:1977tib} which is beyond any experimental reach in the foreseeable future. But many physics beyond the standard model (BSM) scenario, specially new physics related to the generation and smallness of the neutrino mass, can lead to unsuppressed charged LFV processes \cite{Cheng:1977nv, Cheng:1980tp, Lim:1981kv}\footnote{For general condition of tree-level and one-loop lepton flavor violating processes, please see \cite{Blum:2007he}.}  which are within the reach of currently operating and future experiments. For theoretical and experimental status of charged LFV, please see \cite{Kuno:1999jp, Mihara:2013zna, Bernstein:2013hba, Vicente:2015cka, Lindner:2016bgg, Calibbi:2017uvl}.

A well motivated model of neutrino mass generation which addresses the origin of the neutrino mass and the particle nature of the Dark matter (DM) in our universe, is Krauss-Nasri-Trodden (KNT) model \cite{Krauss:2002px} where DM particle radiatively generate the mass of the neutrino at three loops and additional BSM particles having mass at the O(TeV) range, can be accessible to the LHC or future hadron colliders\footnote{For a recent review on radiative generation of neutrino mass, please see \cite{Cai:2017jrq}.}. In \cite{Krauss:2002px}, the additional BSM fields are two charged singlets $S_{1}^{+}$ and $S_{2}^{+}$ and three fermion singlets $N_{R_{1,2,3}}$ which are right handed (RH) neutrinos. A $Z_{2}$ symmetry with action $\{S_{2}^{+},\,N_{R_{i}}\}\rightarrow \{-S_{2}^{+},\,-N_{R_{i}}\}$ is also imposed to  prevent the tree-level Dirac mass of the neutrino after electroweak symmetry breaking, and ensures stability of  the lightest RH neutrino, $N_{R_{1}}$, thereby giving a DM candidate. 

Consequently, the three-loop topology of radiative neutrino mass diagram remains invariant \cite{Chen:2014ska}, if one replaces $S_{2}^{+}$ with larger scalar multiplet, $\mathbf{\Phi}$, which has integer isospin, $j_{\phi}$ and hypercharge\footnote{Here, the electric charge is $Q=T^{3}+Y$.}, $Y_{\phi}=1$ under $SU(2)_{L}\times U(1)_{Y}$ and $N_{R_{i}}$ are replaced with fermionic multiplet $\mathbf{F_{i}}$, $i=1,\,2,\,3$, with integer isospin, $j_{F}$ and hypercharge, $Y_{F}=0$. In this scenario, the DM candidate is the lightest neutral component of $\mathbf{F_{1}}$, i.e. $F^{0}_{1}$. Therefore the immediate generalization of KNT model is \cite{Ahriche:2014cda} where the particle content is taken as, $\mathbf{\Phi}$ with $(j_{\phi},Y_{\phi})=(1,1)$ and $\mathbf{F_{i}}$ with $(j_{F},Y_{F})=(1,0)$. Here, $Z_{2}$ symmetry is still needed to enforce the Dirac mass term of neutrinos to be zero at tree-level. 

In addition, no yukawa terms with SM fermion that give rise to the Dirac neutrino mass, are allowed in the Lagrangian if the KNT particle content is extended with, $\mathbf{\Phi}$ that has $(j_{\phi},Y_{\phi})=(2,1)$ and $\mathbf{F_{i}}$ with $(j_{F},Y_{F})=(2,0)$ to generate the neutrino mass at three-loop level \cite{Ahriche:2014oda}. Therefore, there is no need to use $Z_{2}$ symmetry for that purpose. But the viable dark matter candidate in the model, which is $F_{1}^{0}$ majorana fermion, has one-loop decay process which depends on $\lambda S_{1}^{-}\mathbf{\Phi^{\dagger}}.\mathbf{\Phi}.\mathbf{\Phi}$ term in the scalar potential. From the bound on dark matter mean life-time \cite{Gustafsson:2013gca}, which is of the order $10^{25}-10^{27}$ sec, the $\lambda$ coupling has to be $\lambda\sim 10^{-26}-10^{-27}$ for TeV mass-ranged DM. Moreover, the neutrino sector of the model doesn't depend on this coupling in any way. Therefore in the limit, $\lambda\rightarrow 0$, the softly broken accidental $Z_{2}$ symmetry becomes exact and ensures the stability of the DM.

Consequently, one can go to the next higher scalar and fermion representations in this class of generalized KNT models. In the case of $\mathbf{\Phi}$ with $(j_{\phi},Y_{\phi})=(3,1)$ and $\mathbf{F_{i}}$ with $(j_{F},Y_{F})=(3,0)$, the field content of the model not only prevents the appearance of Dirac mass term for neutrino but also the $\lambda$ term in the scalar potential which would have prevented DM to be absolutely stable \cite{Ahriche:2015wha}. The direct product of two $SU(2)$ scalar representations, $\mathbf{\Phi}\otimes \mathbf{\Phi}$ gives $j_{\phi}\otimes j_{\phi}=\oplus_{J} J\supset j'_{\phi}$ where $j'_{\phi}$ has the same isospin value as $j_{\phi}$ and therefore forms an invariant in the term $\lambda S_{1}^{-}\mathbf{\Phi^{\dagger}}\otimes\mathbf{\Phi}\otimes\mathbf{\Phi}$ but it is either symmetric or antisymmetric representation depending on the even-integer or odd-integer isospin value $j_{\phi}$ respectively. As the antisymmetrized $\mathbf{\Phi}\otimes \mathbf{\Phi}$ representations are identically zero for odd-integer isospin, the $\lambda$ term doesn't appear in the scalar potential at renormalizable level and the DM is stable.

The main motivation of this paper is to carry out a comparative study of charged lepton flavor violating processes in this class of generalized KNT models with singlet, triplet, 5-plet and 7-plet. The most studied charged LFV processes are $\mu\rightarrow e \gamma$, $\mu\rightarrow e e \overline{e}$ and $\mu-e$ conversion in the nuclei. The MEG collaboration has put bound on $\mu\rightarrow e \gamma$ process as $\text{Br}(\mu\rightarrow e \gamma)< 4.2\times 10^{-13}$ at $90\%$ C.L \cite{TheMEG:2016wtm}. In addition, the process $\mu\rightarrow e e \overline{e}$ has current limit as, $\text{Br}(\mu\rightarrow e e \overline{e})<1.0\times 10^{-12}$ ($90\%$ C.L) set by SINDRUM collaboration \cite{Bellgardt:1987du}. Moreover, $\mu-e$ conversion processes in nuclei, $\mu\,\text{Au, Ti}\rightarrow e\,\text{Au, Ti}$ have limits on rates, $\text{CR}(\mu-e,\text{Au})<7\times 10^{-13}$ ($90\%$ C.L) \cite{Bertl:2006up} and $\text{CR}(\mu-e,\text{Ti})<6.1\times 10^{-13}$ ($90\%$ C.L) \cite{Dohmen:1993mp} set by SINDRUM II collaboration. On the other hand, the future reach on $\mu\rightarrow e \gamma$ is, $\text{Br}(\mu\rightarrow e \gamma)< 5.4\times 10^{-14}$ by MEG II experiment, which will start taking data from 2018 \cite{Cattaneo:2017psr}. The Mu3e experiment, which will begin its run on 2019, will have reach $\text{Br}(\mu\rightarrow e e \overline{e})< 10^{-16}$ \cite{Bravar:2017tbs}. In addition, $\mu-e$ conversion experiment Mu2e, which is scheduled to start on 2020, will have $\text{CR}(\mu-e,\text{Al})<6.7\times 10^{-17}$ \cite{theMu2e}. For this reason, we have systematically studied these three processes in each case of Generalized KNT model with respect to the current bounds and future sensitivity limits.

In this article we present the generalized KNT model with larger electroweak multiplets in section \ref{generalizedKNT}. In section \ref{LFVprocesses}, we describe the relevant formulas of charged LFV processes $\mu\rightarrow e \gamma$, $\mu\rightarrow e e \overline{e}$ and $\mu-e$ conversion rate in nuclei in generalized KNT model. Section \ref{result} contains the result of charged LFV processes in this model. We conclude in section \ref{conclusion}. Appendix \ref{loopappendix} contains the loop functions used in calculations of charged LFV processes.

\section{The Model}\label{generalizedKNT}

\subsection{The field content}\label{themodel}

The three-loop radiative neutrino mass generation model contains a charged singlet $S_{1}^{+}\sim (0,0,1)$, a complex scalar multiplet, $\mathbf{\Phi}\sim (0, j_{\phi},1)$ and three real fermion multiplets, $\mathbf{F}_{1,2,3}\sim (0,j_{F},0)$ under $SU(3)_{c}\times SU(2)_{L}\times U(1)_{Y}$. The multiplets are,
\begin{align}
\mathbf{\Phi}&=\left(\phi^{(j_{\phi}+1)},\phi^{(j_{\phi})},...,\phi^{0},...,
\phi^{(-j_{\phi}+2)},\phi^{(-j_{\phi}+1)}\right)^{T}\nonumber\\
\mathbf{F}_{1,2,3}&=\left(F^{(j_{F})},F^{(j_{F}-1)},...,F^{0},...,
F^{(-j_{F}+1)},F^{(-j_{F})}\right)_{1,2,3}^{T}
\label{multiplet}
\end{align}

In this comparative study, we focus on four set of models in this class which we have referred as,
\begin{center}
\begin{tabular}{|c|c|c|}
\hline
Model& $\mathbf{\Phi}$ & $\mathbf{F}_{1,2,3}$\\
\hline
~~Singlet & (0,0,1) & (0,0,0)\\
\hline
~~Triplet & (0,1,1) & (0,1,0)\\
\hline
~~5-plet & (0,2,1) & (0,2,0)\\
\hline
~~7-plet & (0,3,1) & (0,3,0)\\
\hline
\end{tabular}
\end{center}

The SM Lagrangian is extended in the following way,
\begin{equation}
{\cal L}\supset {\cal L}_{SM}+\{f_{\alpha\beta} \overline{L^{c}_{\alpha}}.L_{\beta}S_{1}^{+}
+g_{i \alpha}\overline{\mathbf{F}_{i}}.\mathbf{\Phi}.e_{\alpha_{R}}+h.c\}-\frac{1}{2}\overline{\mathbf{F}^{c}_{i}}M_{F_{ij}}\mathbf{F}_{j}-V(H,\mathbf{\Phi},S_{1})+h.c
\label{eq1}
\end{equation} 
where, c denotes the charge conjugation and dot sign, in shorthand, refers to appropriate $SU(2)$ contractions. Also $L_{\alpha}$ and $e_{R_{\alpha}}$ are the LH lepton doublet and RH charged leptons respectively and Greek alphabet $\alpha$ stands for generation index. Moreover, $[F]_{\alpha\beta}=f_{\alpha\beta}$  and $[G]_{i\alpha}=g_{i\alpha}$ are $3\times 3$ complex antisymmetric and general complex matrices respectively. Finally, $H$ denotes the SM Higgs doublet.

The scalar potential is given by,
\begin{equation}
V(H,\mathbf{\Phi},S_{1})=V(H)+V(\mathbf{\Phi})+V(S_{1})+V_{1}(H,\mathbf{\Phi})+V_{2}(H,S_{1})+V_{3}(\mathbf{\Phi},S_{1})
\label{eq2}
\end{equation}
The three-loop neutrino mass generation depends on the $V_{3}$ term as follows,
\begin{equation}
V_{3}\supset\frac{\lambda_{S}}{4}(S^{-}_{1})^{2}\mathbf{\Phi}.\mathbf{\Phi}+\text{h.c}
\label{eq30}
\end{equation}

\subsection{Mass splittings in the Multiplets}\label{masssplit}

At the tree-level, the components of fermion multiplet, $\mathbf{F}_{i}$ are mass degenerate. Moreover we work in the generation basis where $M_{F_{ij}}=\text{diag}(M_{F_{1}},M_{F_{2}},M_{F_{3}})$. We have also considered the non-degenerate mass for the three fermion multiplets, $M_{F_{1}}<M_{F_{2}}<M_{F_{3}}$. 

Consequently, after electroweak symmetry breaking (EWSB), the radiative corrections, for example, loops involving SM gauge bosons, lift the mass degeneracy in the component fields of the fermion multiplets. In the limit, $M_{F}\gg M_{W}$, the mass splitting between the components of charge $Q$ and $Q'$ is, $M_{Q}-M_{Q'}\sim (Q^{2}-Q^{'2})\Delta$ where, $\Delta\equiv \alpha_{W}\sin^{2}(\theta_{w}/2)M_{W}\sim 166$ MeV \cite{Cirelli:2005uq}.

On the other hand, the component fields of the scalar multiplet, after EWSB, can have splittings at the tree level due to the following term in $V_{1}(H,\mathbf{\Phi})$,
\begin{equation}
V_{1}(H,\mathbf{\Phi})\supset \lambda_{H\phi_{2}}(\mathbf{\Phi}^{*}.H).(H^{*}.\mathbf{\Phi})\label{eq5}
\end{equation}

The maximum splitting among the masses of the component fields in the electroweak multiplet is bounded by the constraints on the Electroweak Precision observables (EWPO) \cite{Peskin:1991sw, Barbieri:2004qk}. Here we consider the constraint on the T parameter as it is the most sensitive EWPO on mass splitting in scalar multiplet or in other words, isospin breaking in the multiplet.
\begin{figure}[h!]
\centerline{\includegraphics[width=11cm]{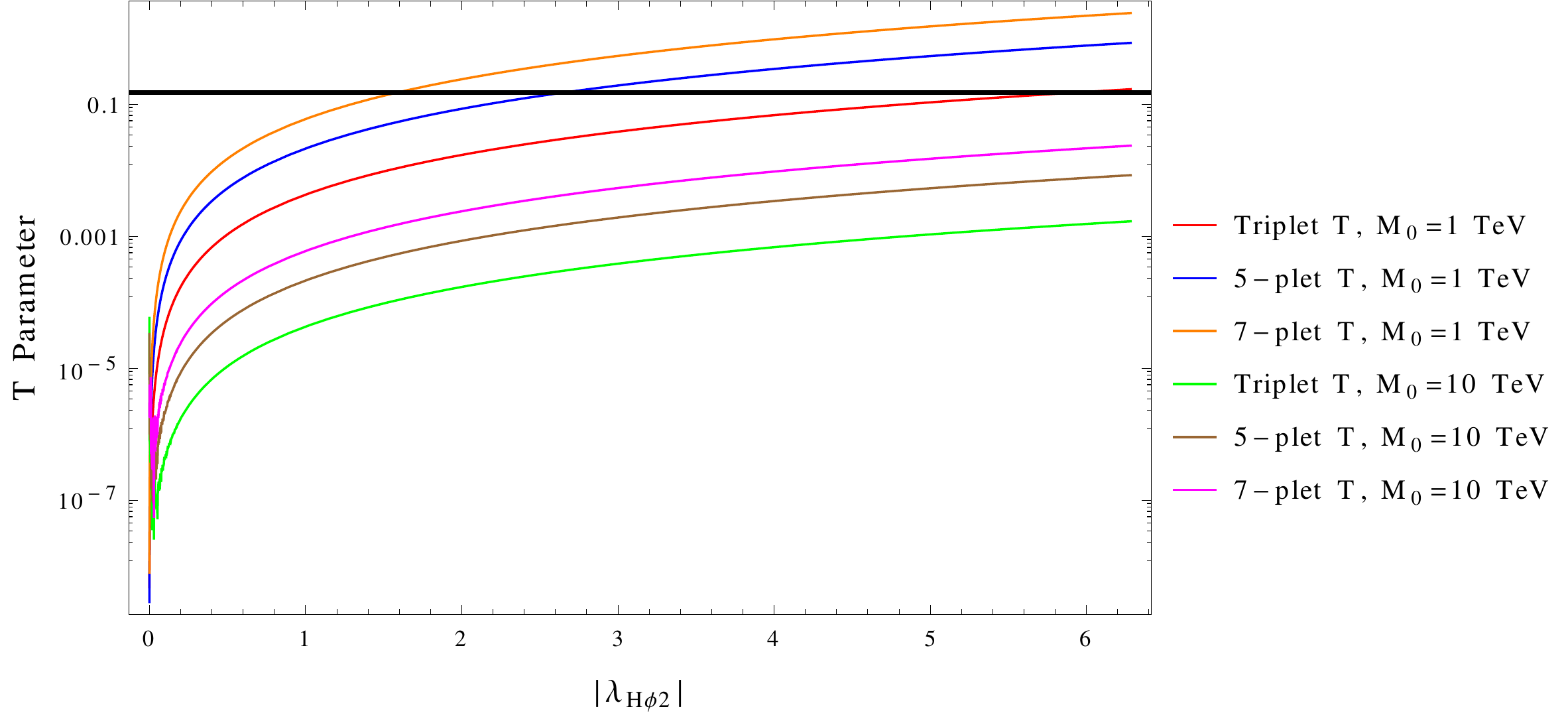}}
\caption{Correlation between $\lambda_{H\phi 2}$ and $T$ parameter. Here we have used two values of $M_{0}$ which is the invariant mass from the Lagrangian, $M_{0}=1$ TeV and $M_{0}=10$ TeV respectively. Also the black line represents the maximum bound on the T parameter, $T=0.07\pm 0.08$ ($68\%$ C.L.) \cite{Patrignani:2016xqp}.}
\label{fig1}
\end{figure}
Therefore, larger value of the coupling, $\lambda_{H\phi 2}$ leads to the larger splitting in the scalar component fields in the multiplet. On the other hand, if $M_{0}=10$ TeV and $\lambda_{H\phi 2}=2\pi$, the splitting between any two components of the scalar multiplet, allowed by the EWPO constraints, is very small as $\Delta m^{2}_{ij}/M^{2}_{0}\sim 10^{-3}$. Again, with $M_{F_{i}}\sim 10$ TeV, the radiative mass splittings between two components of fermionic multiplet leads to $\Delta m^{2}_{F_{ij}}/M^{2}_{0}\sim 10^{-4}$. Therefore, for scalar and fermion multiplets' mass in the TeV range, the mass splittings are numerically negligible therefore we consider this scenario as 'near degenerate' case and make proper approximations in our subsequent analysis.

\subsection{Three-loop radiative neutrino mass}\label{3loopmass}
The neutrino mass is generated radiatively at three loops. In the near degenerate case, we neglect the small mass splittings and have \cite{Krauss:2002px, Ahriche:2014cda, Ahriche:2014oda, Ahriche:2015wha},
\begin{equation}
(M_{\nu})_{\alpha\beta}=\frac{c \lambda_{S}}{(4\pi^{2})^{3}}\frac{m_{\gamma}m_{\delta}}{M_{\phi}}f_{\alpha\gamma}f_{\beta\delta}g^*_{\gamma i}g^*_{\delta i}F\left(\frac{M_{F_{i}}^{2}}{M_{\phi}^{2}},\frac{m_{S}^{2}}{M_{\phi}^{2}}\right)
\label{neumass}
\end{equation}
where, $c=1$, $c=3$, $c=5$ and $c=7$ are for singlet, triplet, 5-plet and 7-plet cases respectively. Eq.(\ref{neumass}) can be written in compact form,
\begin{equation}
M_{\nu}=X.\Lambda.X^{T},\,\,\,\text{with}\,\,\,X=FM_{l}G^{\dagger}
\label{neumasscompact}
\end{equation}
Here, $M_{l}$ is the diagonal charged lepton mass matrix and $\Lambda=\text{diag}(\Lambda_{1},\Lambda_{2},\Lambda_{3})$, where $\Lambda_{i}$ is associated with $\mathbf{F}_{i}$.

\begin{figure}[h!]
\centerline{\includegraphics[width=10cm]{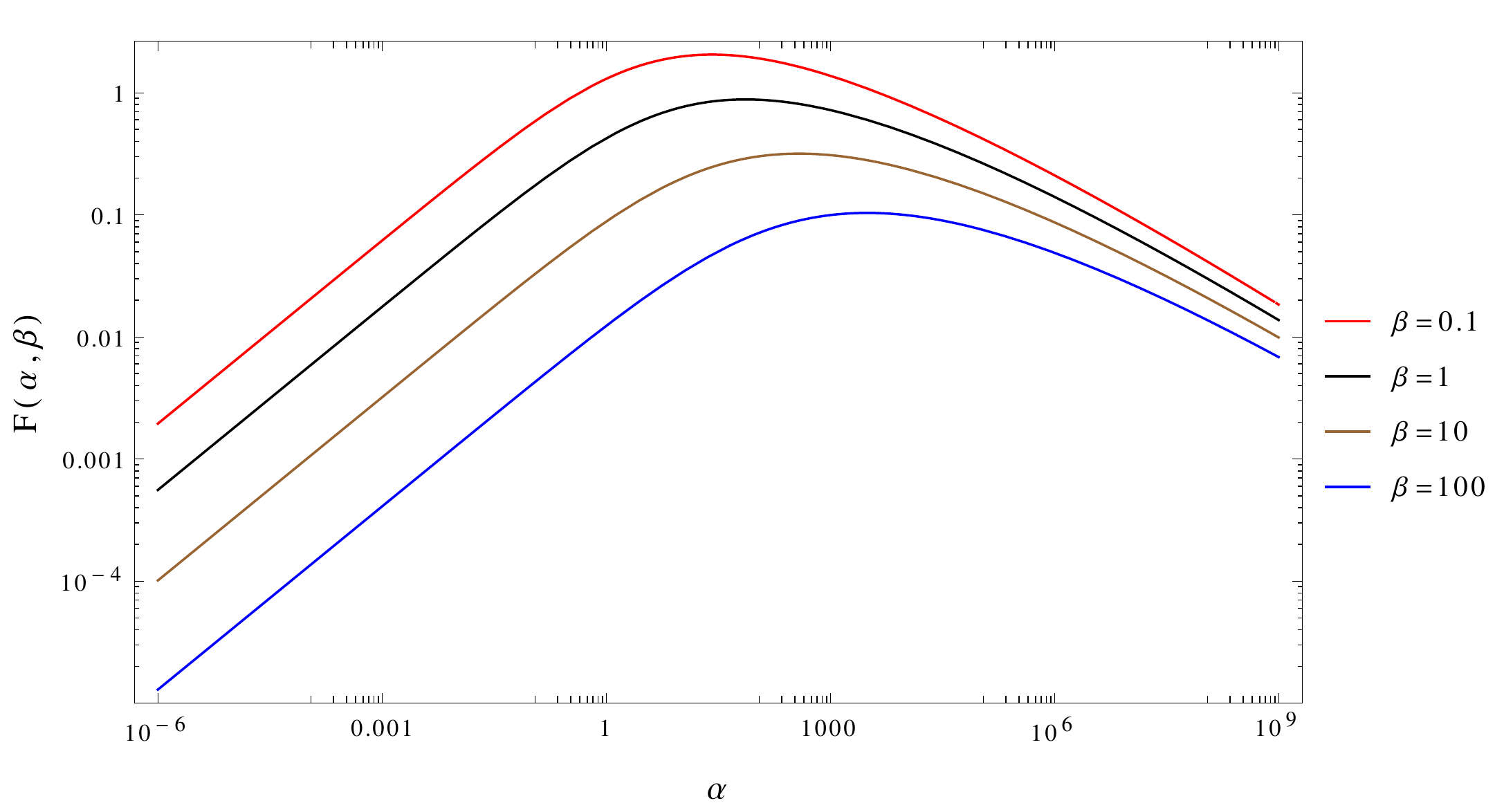}}
\caption{The function $F(\alpha,\beta)$.}
\label{Ffunction}
\end{figure}

The loop function $F$ with $\alpha=M_{F_{i}}^{2}/M_{\phi}^{2}$ and $\beta=m_{S}^{2}/M_{\phi}^{2}$, is given by,
\begin{equation}
F(\alpha,\beta)=\frac{\sqrt{\alpha}}{8\beta^{2}}\int_{0}^{\infty}dr\frac{r}{r+\alpha}I(r,\beta)^{2}
\label{loopeq1}
\end{equation}
and the function $I(r,\beta)$ is
\begin{equation}
I(r,\beta)=\ln[r(\eta_{+}-1)(1-\eta_{-})]-\eta_{+}\ln\left[\frac{\eta_{+}-1}{\eta_{+}}\right]-\eta_{-}\ln\left[\frac{\eta_{-}-1}{\eta_{-}}\right]-\frac{1+r}{r}\ln[1+r]
\label{loopeq2}
\end{equation}
where
\begin{equation}
\eta_{\pm}(r,\beta)=\frac{1}{2r}\left(1+r-\beta\pm\sqrt{(1+r-\beta)^2+4 r \beta}\right)
\label{loopeq3}
\end{equation}
The behavior of function $F(\alpha,\beta)$ with $\alpha$ and $\beta$ is shown in Fig. \ref{Ffunction}.

The neutrino mass matrix, $M_{\nu}$ of Eq.(\ref{neumass}), can be diagonalized as
\begin{equation}
M_{\nu}=U_{\text{PMNS}}.\hat{m}_{\nu}.U^{T}_{\text{PMNS}}
\label{nudiag}
\end{equation}
where, $\hat{m}_{\nu}=\text{diag}(m_{\nu_{1}},m_{\nu_{2}},m_{\nu_{3}})$ and it
contains 7 independent parameters which are two masses $m_{\nu_{2,3}}$ that can be determined assuming either normal or inverted hierarchy by using experimentally measured \cite{Patrignani:2016xqp} two mass squared differences $\Delta m^{2}_{\text{atm}}$ and $\Delta m^{2}_{\text{solar}}$, three mixing angles $\theta_{12}$, $\theta_{23}$ and $\theta_{13}$ and still to be determined, one Dirac phase $\delta_{\text{CP}}$ and one Majorana phase $\alpha_{M}$ in $U_{\text{PMNS}}$ matrix. Here due to $\text{det}\,F=0$, the lowest neutrino mass is $m_{\nu_{1}}=0$ and it also implies one Majorana phase of $U_{\text{PMNS}}$ to be zero. On the other hand, the matrices $F$ contains six and $G$ contains 18 degrees of freedom. As there is no one-to-one correspondence between low energy neutrino parameters in $M_{\nu}$ and the parameters of $F$, $G$ and $\Lambda$, we numerically determine the set $\{f_{\alpha\beta},\,g_{i\alpha},\, M_{F_{1,2,3}},\,M_{\phi},\,m_{S},\,\lambda_{S}\}$ which satisfy the following relation,
\begin{equation}
\text{Tr}(M^{\dagger}_{\nu}M_{\nu})=\text{Tr}(X^{*}\,\Lambda\,X^{\dagger}\,X\,\Lambda\,X^{T})
\label{neueq}
\end{equation}
We have used this relation because there are no unique $F$ and $G$ which satisfy the low energy neutrino constraints, $U_{\text{PMNS}}$. Therefore one can always find another set of $F'$ and $G'$ through orthogonal transformation, $F'\rightarrow VFV^{T}$ and bi-unitary transformation, $G'\rightarrow W G Y^{\dagger}$ where, $V$, $W$ and $Y$ are unitary matrices, to satisfy the low energy constraints.

\section{Charged Lepton Flavor Violating Processes}\label{LFVprocesses}
As the charged LFV processes, $\mu\rightarrow e\gamma$, $\mu\rightarrow e e \overline{e}$ and $\mu-e$ conversion in Au and Ti nuclei have the most stringent experimental constraints, we focus our study on these three LFV processes in generalized KNT model with singlet, triplet, 5-plet and 7-plet respectively. 

\subsection{$\mu\rightarrow e\,\gamma$}\label{mutoegamma}

The branching ratio for $\mu\rightarrow e\gamma$, normalized by 
$\text{Br}(\mu\rightarrow e\overline{\nu_{e}}\nu_{\mu})$, is 
\begin{equation}
 \text{Br}(\mu\rightarrow e\gamma)=\frac{3(4\pi)^3\alpha_{em}}{4G_{F}^2}|A_{D}|^2\,\text{Br}(\mu\rightarrow e\nu_{\mu}\overline{\nu_{e}})
 \label{mutoegma}
\end{equation}
where
\begin{equation}
 A_{D}=A^{(1)}_{D}+A^{(2)}_{D}+A^{(3)}_{D}
 \label{dipole1}
\end{equation}
where
\begin{eqnarray}
 A^{(1)}_{D}&=&\sum_{i=1}^{3}\sum_{\phi}\frac{g^{*}_{e i}g_{i\mu}q_{\phi}}{32\pi^2}\frac{1}{m^2_{\phi}}F_{1}(x^{q}_{i\phi})\\ 
 %\label{dipole2}
A^{(2)}_{D} &=&-\sum_{i=1}^{3}\sum_{F_{i}}\frac{g^{*}_{e i}g_{i\mu}q_{F_{i}}}{32\pi^2}\frac{1}{m^2_{\phi}}F_{2}(x^{q}_{i\phi})\\
 %\label{dipole3}
 A^{(3)}_{D}&=&\frac{f^{*}_{e \tau}f_{\tau \mu}}{192 \pi^2}\frac{1}{m_{S}^{2}}
%\label{dipole4}
\end{eqnarray}
where $m_{\phi}$ and $q_{\phi}$ ($q_{F}$ ) are the corresponding mass and the electric charge respectively of the scalar (fermion) component $\phi^{(q)}$ ($F^{(q)}_{i}$), $x^{q}_{i\phi}=m^{2}_{F^{(q-1)}_{i}}/m^{2}_{\phi^{(-q)}}$ and $\phi$ ($F_{i}$) runs over all the charged  components of the  scalar (fermion) multiplet $\mathbf{\Phi}$ ($\mathbf{F}_{i}$).  Note that $A^{(1)}_{D}$ and $A^{(2)}_{D}$ involve RH charged leptons whereas, $A^{(3)}_{D}$ involves LH charged leptons.

\begin{figure}[h!]
\centerline{\includegraphics[width=10cm]{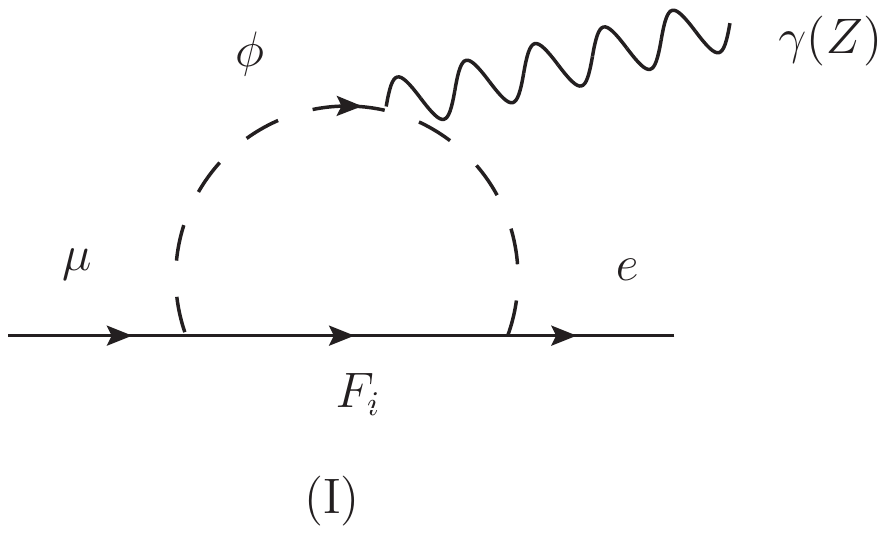}\hspace{-6cm}
\includegraphics[width=10cm]{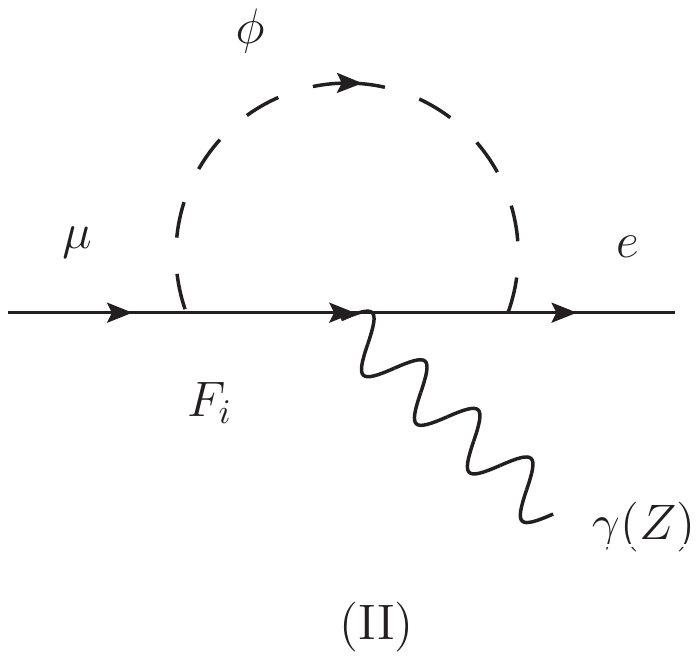}\hspace{-6cm}
\includegraphics[width=10cm]{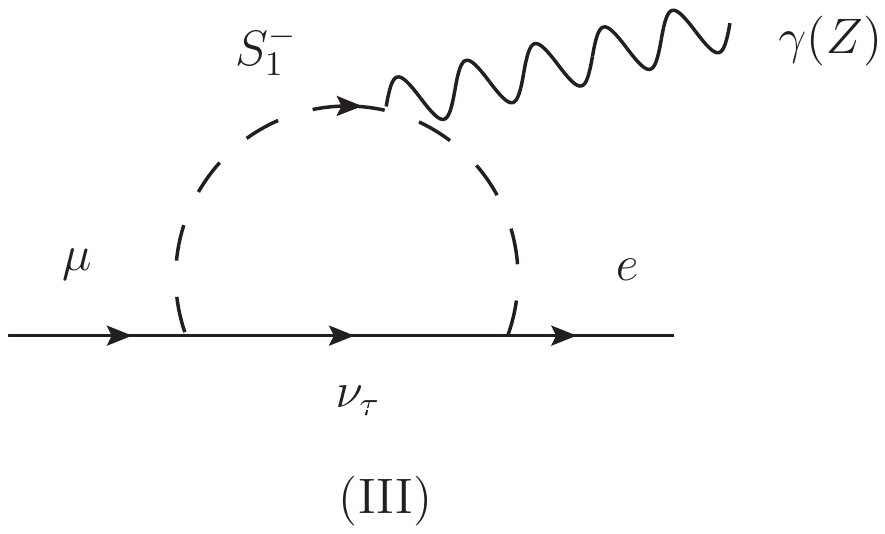}\hspace{-6cm}
\includegraphics[width=10cm]{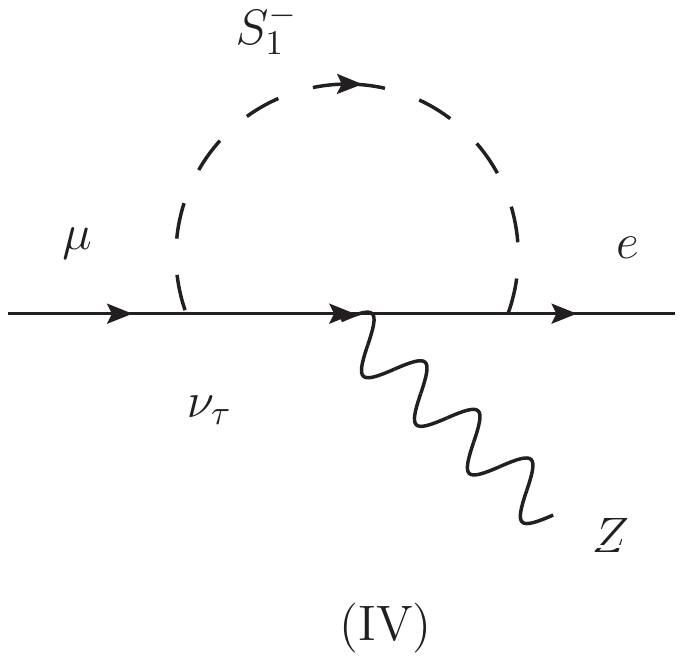}}
\vspace{-9.5cm}
\caption{One loop diagrams contributing to effective $\mu e \gamma$ and $\mu e Z$ vertices. For simplicity, we have not included self energy diagrams where $\gamma (Z)$ line is attached to external fermions.}
\label{muegammaver}
\end{figure}

Fig. \ref{muegammaver} (I) and (II), where external $\gamma$ line is attached to charged scalars and charged fermions respectively and give dipole contributions $A^{(1)}_{D}$ and $A^{(2)}_{D}$ that come from pairs, $(\phi^{(-q)},F_{i}^{(q-1)})$ where, $q=-j_{\phi}+1,...,j_{\phi}+1$. Since the mass splittings among the component fields of both the scalar and fermion multiplet are small as pointed out in section \ref{masssplit}, we can consider the near degenerate limit and in this case, there are cancellations in $A^{(2)}_{D}$ because degenerate fermion components of opposite electric charge have photon line attached to it and therefore sum over all fermion components renders it to $A^{(2)}\sim 0$. Moreover the same cancellations take place in $A^{(1)}_{D}$ when scalar components of opposite electric charge have photon line attached to it. Therefore in the case of triplet we have non-negligible contributions from $(\phi^{--},F_{i}^{+})$ and $(\phi^{-},F_{i}^{0})$ pairs in $A^{(1)}_{D}$. For 5-plet, we have non-negligible contributions in $A^{(1)}_{D}$ from $(\phi^{---},F_{i}^{++})$ and $(\phi^{--},F^{+}_{i})$ pairs. Finally, for 7-plet, the only non-negligible contributions in $A^{(1)}_{D}$ come from $(\phi^{----},F^{+++}_{i})$ and $(\phi^{---},F_{i}^{++})$ pairs. On the other hand, singlet case only involves $(\phi^{-},F_{i}^{0})$ pair as the non-zero contribution to $A^{(1)}_{D}$. On the other hand, Figure \ref{muegammaver} (III) gives dipole contribution $A^{(3)}_{D}$ coming from $(S^{-}_{1},\nu_{\tau})$ pair for all cases. 

\subsection{$\mu\rightarrow ee\overline{e}$}\label{mueeesection}

In the generalized KNT model, the 3-body lepton flavor violating decay mode $\mu\rightarrow ee\overline{e}$ receives the contributions from $\gamma$-penguin diagrams, Z-penguin diagrams, and Box diagrams. In this model, Higgs penguin diagram doesn't contribute to this process. Therefore, the branching ratio of $\mu\rightarrow e e \overline{e}$ is given
\cite{Hisano:1995cp, Arganda:2005ji, Abada:2014kba, Arganda:2014lya} as
\begin{align}
 \text{Br}(\mu\rightarrow ee\overline{e})&=\frac{3(4\pi)^2\alpha_{em}^2}{8G_{F}^2}
 \left[|A_{ND}|^2+|A_{D}|^2\left(\frac{16}{3}\text{ln}\frac{m_{\mu}}{m_{e}}-\frac{22}{3}\right)+\frac{1}{6}|B|^2\right.\notag\\
 &+\left.\frac{1}{3}(2|F^{L}_{Z}|^2+|F^{R}_{Z}|^2)+\left(-2A_{ND}A^*_{D}+\frac{1}{3}A_{ND}B^*-\frac{2}{3}A_{D}B^*+\text{h.c}\right)\right]\notag\\
 &\times \text{Br}(\mu\rightarrow e\overline{\nu_{e}}\nu_{\mu})
 \label{mutoeee}
\end{align}
where $A_{D}$ and $A_{ND}$ are the dipole and 
non-dipole contributions from the photonic penguin diagrams respectively. $F^{L}_{Z}$ and $F^{R}_{Z}$ are given as
\begin{equation}
 F^{L}_{Z}=\frac{F_{Z}g^{l}_{L}}{g^2m_{Z}^2\sin^2\theta_{W}}\,\,\,,\,\,\,F^{R}_{Z}=\frac{F_{Z}g^{l}_{R}}{g^2m_{Z}^2\sin^2\theta_{W}}
 \label{zcontrib}
\end{equation}
Here, $F_{Z}$ is the Z-penguin contribution and $g^{l}_{L}$ and $g^{l}_{R}$ are the Z-boson coupling to the left-handed (LH) and right-handed (RH) charged leptons respectively. Also
$B$ represents the contribution from the box diagrams.

\subsubsection{$\gamma$-penguin contribution}\label{gammapenguin}
The $\gamma$ penguin diagram can be obtained by attaching $e-\overline{e}$ fermion line to $\gamma$ line in Fig. \ref{muegammaver} (I), (II) and (III). The dipole contribution of $\gamma$-penguin diagrams are same as in section \ref{mutoegamma}. So we consider here the non-dipole contribution which is,
\begin{equation}
A_{ND}=A^{(1)}_{ND}+A^{(2)}_{ND}+A^{(3)}_{ND}
\label{nondipole1}
\end{equation}
Here,
\begin{align}
 A^{(1)}_{ND}&=\sum_{i=1}^{3}\sum_{\phi}\frac{g^{*}_{e i}g_{i\mu}q_{\phi}}{32\pi^2}\frac{1}{m^2_{\phi}}G_{1}(x^{q}_{i\phi})
 \label{nondipole2}\\
 A^{(2)}_{ND}&=-\sum_{i=1}^{3}\sum_{F_{i}}\frac{g^{*}_{e i}g_{i\mu}q_{F_{i}}}{32\pi^2}\frac{1}{m^2_{\phi}}G_{2}(x^{q}_{i\phi})
 \label{nondipole3}\\
A^{(3)}_{ND}&=\frac{f^{*}_{e \tau}f_{\tau \mu}}{288 \pi^2}\frac{1}{m_{S}^{2}}
\label{nondipole4}
\end{align}
 
The non-dipole contributions $A^{(1)}_{ND}$, $A^{(2)}_{ND}$ and $A^{(3)}_{ND}$ are associated with Fig. \ref{muegammaver} (I), (II) and (III) respectively with $\gamma$ line having $e-\overline{e}$ fermion line attached to it. The loop functions $G_{1}(x)$ and $G_{2}(x)$ are given in appendix \ref{loopappendix}. As in the case of dipole contributions, in the near degenerate mass limit, we again have cancellations among the charged fermions with opposite electric charge in $A^{(2)}_{ND}$ therefore, $A^{(2)}_{ND}\sim 0$. In addition, the contributions from charged scalar with opposite electric charge get canceled in $A^{(1)}_{ND}$. Once more the non-negligible contributions in $A^{(1)}_{ND}$ come from $(\phi^{--},F_{i}^{+})$ and $(\phi^{-},F_{i}^{0})$ in the case of triplet, $(\phi^{---},F_{i}^{++})$ and $(\phi^{--},F^{+}_{i})$ pairs for the case of 5-plet and $(\phi^{----},F^{+++}_{i})$ and $(\phi^{---},F_{i}^{++})$ pairs for the case of 7-plet respectively.

\subsubsection{Z-Penguin Contribution}\label{zpenguin}

The Z-penguin diagram can be obtained from Fig \ref{muegammaver} (I)-(IV) by attaching $e-\overline{e}$ fermion line attaching to Z boson line. Its contribution to $\mu\rightarrow ee\overline{e}$ can be arranged into two parts,
\begin{equation}
F_{Z}=F^{(1)}_{Z}+F^{(2)}_{Z}
\label{zpenguin1}
\end{equation}
where $F^{(1)}_{Z}$ is the contribution associated with Fig. \ref{muegammaver} (I) and (II) with Z line. On the other hand, $F^{(2)}_{Z}$ is the contribution associated with Fig. \ref{muegammaver} (III) (with Z line) and (IV). They are given as,
\begin{align}
 F^{(1)}_{Z}& =-\frac{1}{16\pi^2}\sum_{i=1}^{3}\sum_{(\phi,F_{i})}\left\{g^{*}_{e i}g_{i\mu}\,g_{ZF_{i}\overline{F_{i}}}\,%
 \left[\left(2C_{24}(m_{\phi},m_{F_{i}},m_{F_{i}})+\frac{1}{2}\right)+ m^{2}_{F_{i}} C_{0}(m_{\phi},m_{F_{i}},m_{F_{i}})\right]\right. \notag \\
 &\left. +2\,g^{*}_{e i}g_{i\mu}\,g_{Z\phi}\,%
 C_{24}(m_{F_{i}},m_{\phi},m_{\phi})+ g^{*}_{e i}g_{i\mu}g^{l}_{R}B_{1}(m_{F_{i}},m_{\phi})\right\}\label{zpenguin2}\\
 F^{(2)}_{Z}&=-\frac{1}{16\pi^2}f^{*}_{e \tau}f_{\tau\mu}\left\{g_{Z\nu\overline{\nu}}\,%
 \left(2C_{24}(m_{S_{1}},0,0)+\frac{1}{2}\right)+2g_{ZS_{1}}C_{24}(0,m_{S_{1}},m_{S_{1}})\right.\notag\\ 
 & \left. +g^{l}_{L}B_{1}(0,m_{S_{1}})\right\}\label{zpenguin3}
 \end{align}
where the sum over pairs $(\phi,F_{i})$ implies the pairs of component fields from fermion and scalar multiplet entering into the one-loop process. $g{ZF_{i}\overline{F_{i}}}$, $g_{Z\phi}$, $g_{Z\nu\overline{\nu}}$ and $g_{ZS_{1}}$ are the Z coupling of charged fermion components of $\mathbf{F_{i}}$, scalar components of $\mathbf{\Phi}$, tau neutrino and charged scalar $S_{1}$ respectively. Moreover, $g^{l}_{L}$ and $g^{l}_{R}$ are the Z coupling of the left handed and right handed charged leptons respectively.

In the near degenerate limit, for the triplet, the contributions from following pairs in Eq. (\ref{zpenguin2}) are,
\begin{equation}
F^{(1)}_{Z}(\phi^{--},F^{+}_{i})=-F^{(1)}_{Z}(\phi^{0},F^{-}_{i})
\label{zpenguin40}
\end{equation}

For the 5-plet, the contribution in Eq.(\ref{zpenguin2}) from the following pairs are
\begin{equation}
F^{(1)}_{Z}(\phi^{---},F^{++}_{i})=-F^{(1)}_{Z}(\phi'^{+},F^{--}_{i})\,\,\,\text{and}\,\,\,
F^{(1)}_{Z}(\phi^{--},F^{+}_{i})=-F^{(1)}_{Z}(\phi^{0},F^{-}_{i})
\label{zpenguin4}
\end{equation}
whereas, for 7-plet, the contribution from the following pairs are,
\begin{align}
&\,\,F^{(1)}_{Z}(\phi^{----},F^{+++}_{i})=-F^{(1)}_{Z}(\phi''^{++},F^{---}_{i}),\,\,
F^{(1)}_{Z}(\phi^{---},F^{++}_{i})=-F^{(1)}_{Z}(\phi'^{+},F^{--}_{i})\notag\,\,\,\text{and}\\
&\,\,F^{(1)}_{Z}(\phi^{--},F^{+}_{i})=-F^{(1)}_{Z}(\phi^{0},F^{-}_{i})
\label{zpenguin5}
\end{align}

For singlet, there is only one contribution in $F^{(1)}_{Z}$ which is coming from $(\phi^{-},F_{i}^{0})$ pair. Therefore in all cases, the only non-zero contribution in $F^{(1)}_{Z}$ comes from $(\phi^{-},F^{0}_{i})$ pair.

In addition, $F^{(2)}_{Z}$ in Eq.(\ref{zpenguin3}) is zero because the loop functions satisfy the relation, 
\begin{equation}
2 C_{24}(m_{S_{1}}^{2},0,0)+\frac{1}{2}=2 C_{24}(0,m_{S_{1}}^{2},m_{S_{1}}^{2})=B_{1}(0,m_{S_{1}}^{2})
\label{zpenguin55}
\end{equation}
And Z-couplings are $g_{Z\nu\overline{\nu}}=\frac{g}{2\cos\theta_{W}}$, $g_{ZS_{1}}=-\frac{g\sin^{2}\theta_{W}}{\cos\theta_{W}}$ and $g^{l}_{L}=\frac{g}{\cos\theta_{W}}\left(-\frac{1}{2}+\sin^{2}\theta_{W}\right)$. Therefore the total sum turns out to zero.

\begin{figure}[h!]
\vspace{-1.5cm}
\hspace{-6cm}\centerline{\includegraphics[width=11cm]{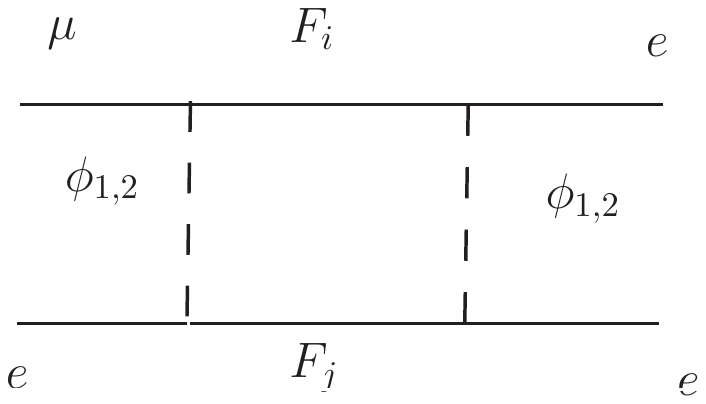}\hspace{-6cm}
\includegraphics[width=11cm]{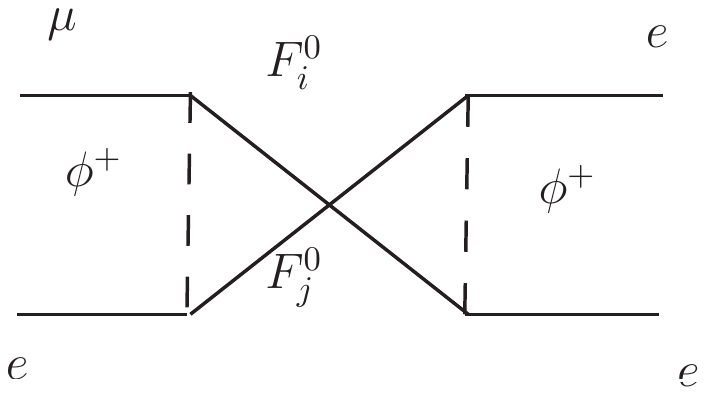}\hspace{-6cm}
\includegraphics[width=11cm]{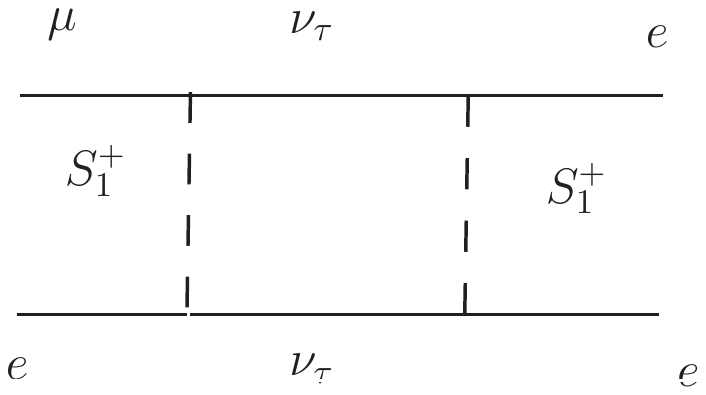}\hspace{-6cm}}
\vspace{-12cm}
\caption{One-loop box topologies associated to Feynman diagrams contributing to $\mu\rightarrow e e \overline{e}$ process.}
\label{boxdig}
\end{figure}

\subsubsection{Box Contribution}\label{boxdiagrams}

The box contribution can be arranged into three parts,
\begin{equation}
B=B^{(1)}+B^{(2)}+B^{(3)}
\label{box1}
\end{equation}
Here, $B^{(1)}$ is the contribution associated with neutral fermion in the loop and involves combination of one-loop box topologies Fig. \ref{boxdig} (left and center). It is given as,
\begin{equation}
 e^2\,B^{(1)}=\frac{1}{16\pi^2}\sum_{i,j=1}^{3}\left[\frac{\tilde{D}_{0}}{2}g^{*}_{ei}g_{i\mu}g^{*}_{ej}g_{je}
 +D_{0}m_{F^{0}_{i}}m_{F^{0}_{j}}g^{*}_{ei}g^{*}_{ei}g_{j\mu}g_{je}\right]
\label{box2}
\end{equation}
where, $\tilde{D}_{0}=\tilde{D}_{0}(m_{F^{0}_{i}},m_{F^{0}_{j}},m_{\phi^{+}},m_{\phi^{+}})$ and
$D_{0}=D_{0}(m_{F^{0}_{i}},m_{F^{0}_{j}},m_{\phi^{+}},m_{\phi^{+}})$.

On the contrary, the charged fermions running in the loop give contribution to $B^{(2)}$ and are associated to Fig. \ref{boxdig} (left). It is given as,
\begin{equation}
 e^2\,B^{(2)}=\frac{1}{32\pi^2}\sum_{i,j=1}^{3}\sum_{F}\sum_{\phi_{1},\phi_{2}}\tilde{D}_{0}(m_{F_{i}},m_{F_{j}},m_{\phi_{1}},m_{\phi_{2}})g^{*}_{ei}g_{i\mu}g^{*}_{ej}g_{je}
\label{box3}
\end{equation}
The sum index $F$ ranges over the charged components of fermion multiplets and $\phi_{1,2}$ indices range over the corresponding scalar components set by the $G$ yukawa term in Eq.(\ref{eq1}). 

For the singlet, the only contribution is $B^{(1)}$ that comes box diagram that involves $\phi^{+}$ and neutral fermion $F^{0}_{i}$. For larger scalar and fermion multiplets, apart from  $B^{(1)}$ contribution also $B^{(2)}$ has contributions from charged fermions as follows. For triplet, box contribution $B^{(2)}$ involves $\phi_{1,2}\in\{\phi^{++},\phi^{0}\}$ for $F=F_{i}^{+}$. For 5-plet, $\phi_{1,2}\in\{\phi^{+++},\phi'^{-}\}$ for $F=F_{i}^{++}$ and $\phi_{1,2}\in\{\phi^{++},\phi^{0}\}$ for $F=F_{i}^{+}$. On the other hand, for 7-plet, $\phi_{1,2}\in\{\phi^{++++},\phi''^{--}\}$ for $F=F_{i}^{+++}$, $\phi_{1,2}\in\{\phi^{+++},\phi'^{-}\}$ for $F=F_{i}^{++}$ and $\phi_{1,2}\in\{\phi^{++},\phi^{0}\}$ for $F=F_{i}^{+}$.

There is also box contribution coming from the charged scalar $S^{+}_{1}$ (Fig. \ref{boxdig} (right)) which is given as
\begin{equation}
e^2 B^{(3)}=-\frac{1}{32\pi^2 m^{2}_{S}}f^{*}_{e\tau}f_{\tau\mu}f^{*}_{e\tau}f_{\tau e}
\label{box4}
\end{equation}

\subsection{$\mu$ to $e$ Conversion Rate}\label{mueconversion}

The conversion rate, normalized by the muon capture rate is 
\cite{Kitano:2002mt, Arganda:2007jw, Crivellin:2014cta}
\begin{align}
 \text{CR}(\mu-e,\text{Nucleus})&=\frac{p_{e}E_{e}m^3_{\mu}G_{F}^2\alpha_{em}^3Z_{eff}^{4}F_{p}^2}{8\pi^2 Z\,\Gamma_{\text{capt}}}
 \left\{|(Z+N)(g^{(0)}_{LV}+g^{(0)}_{LS})+(Z-N)(g^{(1)}_{LV}+g^{(1)}_{LS})|^{2}\right.\notag\\
 &+\left.|(Z+N)(g^{(0)}_{RV}+g^{(0)}_{RS})+(Z-N)(g^{(1)}_{RV}+g^{(1)}_{RS})|^{2}\right\}
 \label{mueconv1}
\end{align}
Here, $Z$ and $N$ are the number of protons and neutrons in the nucleus, $Z_{eff}$ is the effective atomic
charge, $F_{p}$ is the nuclear matrix element and $\Gamma_{\text{capt}}$ represents
the total muon capture rate. $p_{e}$ and $E_{e}$ are the momentum and energy of the electron which is taken as 
$\sim m_{\mu}$. $g^{(0)}_{XK}$ and $g^{(1)}_{XK}$ ($X=L,R$ and $K=V,S$)
in the above expression are given as
\begin{eqnarray}
 g^{(0)}_{XK}=\frac{1}{2}\sum_{q=u,d,s}(g_{XK(q)}G^{(q,p)}_{K}+g_{XK(q)}G^{(q,n)}_{K})\nonumber\\
 g^{(1)}_{XK}=\frac{1}{2}\sum_{q=u,d,s}(g_{XK(q)}G^{(q,p)}_{K}-g_{XK(q)}G^{(q,n)}_{K})
 \label{mueconv2}
\end{eqnarray}
$g_{XK(q)}$ are the couplings in the effective Lagrangian describing $\mu-e$ conversion,
\begin{equation}
 {\cal L}_{eff}=-\frac{G_{F}}{\sqrt{2}}\sum_{q}\left\{[g_{LS(q)}\overline{e}_{L}\mu_{R}+g_{RS(q)}\overline{e}_{R}\mu_{L}]\overline{q}q+
 [g_{LV(q)}\overline{e}_{L}\gamma^{\mu}\mu_{L}+g_{RV(q)}\overline{e}_{R}\gamma^{\mu}\mu_{R}]\overline{q}\gamma_{\mu}q\right\}
\end{equation}
$G^{(q,p)},\, G^{(q,n)}$ are the numerical factors that arise when quark matrix elements
are replaced by the nucleon matrix elements,
\begin{equation}
 \langle p|\overline{q}\Gamma_{K}q|p\rangle=G^{(q,p)}_{K}\overline{p}\Gamma_{K}p\,\,,\,\,
 \langle n|\overline{q}\Gamma_{K}q|n\rangle=G^{(q,n)}_{K}\overline{n}\Gamma_{K}n
 \label{mueconv3}
\end{equation}
For the generalized KNT model, the $\mu-e$ conversion rate receives the $\gamma$ and Z penguin
contributions where the quark line is attached to photon and Z-boson lines in the respective penguin diagrams. It also doesn't receive any box contribution because there is no coupling
between $\mathbf{\Phi}$ and quarks. The relevant effective coupling for the conversion
in this model is
\begin{eqnarray}
 g_{LV(q)}&=&g^{\gamma}_{LV(q)}+g^{Z}_{LV(q)}\nonumber\\
 g_{RV(q)}&=&g_{LV(q)}|_{L\leftrightarrow R}\nonumber\\
 g_{LS(q)}&\approx& 0\,\,\,,\,\,\,g_{RS(q)}\approx 0\nonumber
\end{eqnarray}
The relevant couplings are
\begin{eqnarray}
 g^{\gamma}_{RV(q)}&=&\frac{\sqrt{2}} {G_{F}}e^2Q_{q}\left[(A^{(1)}_{ND}+A^{(2)}_{ND})-(A^{(1)}_{D}+A^{(2)}_{D})\right],\,\,
 g^{\gamma}_{LV(q)}=\frac{\sqrt{2}}{G_{F}}e^2Q_{q}(A^{(3)}_{ND}-A^{(3)}_{D})\nonumber\\
 g^{Z}_{RV(q)}&=&-\frac{\sqrt{2}}{G_{F}}\frac{g^{q}_{L}+g^{q}_{R}}{2}\frac{F^{(1)}_{Z}}{m_{Z}^2},\,\,
 g^{Z}_{LV(q)}=-\frac{\sqrt{2}}{G_{F}}\frac{g^{q}_{L}+g^{q}_{R}}{2}\frac{F^{(2)}_{Z}}{m_{Z}^2}
 \label{mueconv5}
\end{eqnarray}
Here $Q_{q}$ is the electric charge of the quarks and Z boson couplings to the quarks are
\begin{equation}
 g^{q}_{L}=\frac{g}{\cos\theta_{W}}(T^{q}_{3}-Q_{q}\sin^2\theta_{W})\,\,,\,\,
 g^{q}_{R}=-\frac{g}{\cos\theta_{W}}Q_{q}\sin^2\theta_{W}
 \label{mueconv6}
\end{equation}
Also the relevant numerical factors for nucleon matrix elements are
\begin{equation}
 G^{(u,p)}_{V}=G^{(d,n)}_{V}=2\,\,,\,\,G^{(d,p)}_{V}=G^{(u,n)}_{V}=1
 \label{mueconv7}
\end{equation}

In the near degenerate limit, there will be cancellation in $A_{D}$, $A_{ND}$ and $F_{Z}$ contributions for triplet, 5-plet and 7-plet cases as pointed out in sections \ref{mutoegamma}, \ref{gammapenguin} and \ref{zpenguin}. Therefore, $\mu-e$ conversion rate will be also suppressed compared to the $\mu\rightarrow e e \overline{e}$ process in the KNT model.

\section{Result and Discussion}\label{result}

\subsection{Viable Parameter Space}\label{viableparamatersp}

The parameter space of generalized KNT model for singlet, triplet, 5-plet and 7-plet in the near degenerate limit is taken as $\{f_{\alpha\beta},\,g_{i\alpha},\, M_{F_{1,2,3}},\,M_{\phi},\,m_{S},\,\lambda_{S}\}$ which enter into the neutrino mass generation in Eq.(\ref{neumass}). 

Here we briefly present the collider constraints and future reach on the masses of the fermion and scalar multiplets in this model. The sensitivity study \cite{Ahriche:2014xra} on the process $e^{+} e^{-}\rightarrow S^{+}_{1} S^{-}_{1}\rightarrow l^{+}_{\alpha} l^{-}_{\beta}+E_{\text{miss}}$ in KNT model at future International Linear Collider (ILC) with $\sqrt{s}=1$ TeV showed that $m_{S}\simgt 240$ GeV. On the other hand, it was shown in \cite{Cherigui:2016tbm} that for tri-lepton final states via $pp\rightarrow l^{\pm}l^{\pm}S^{-*}_{1}\rightarrow l^{\pm}l^{\pm}l^{\pm}+E_{\text{miss}}$ at LHC with $\sqrt{s}=14$ TeV and luminosity, $L=300\,\text{fb}^{-1}$, the discovery reach for $S^{+}_{1}$ increases up to $m_{S}\simlt 4$ TeV.

In addition, we have $F^{0}_{1}$ to be DM candidate and that sets $M_{\phi}>M_{F_{1}}$. Based on searches of disappearing track signatures from long-lived charginos that is nearly mass-degenerate with a neutralino at LHC with $\sqrt{s}=14 $ TeV and $L=36.1\,\text{fb}^{-1}$ \cite{Aaboud:2017mpt}, we can re-interpret the exclusion limits for fermion components as $m_{F^{\pm}_{1}}\simgt 600$ GeV for lifetime,$\tau_{F^{\pm}_{1}}=1$ ns. Moreover, for wino-like minimal DM models that resembles fermion multiplets of KNT model, future collider with $\sqrt{s}=100 $ TeV and $L=3\,\text{ab}^{-1}$ \cite{Cirelli:2014dsa}, will improve this limit to $m_{F^{0}_{1}}\simgt 3.2$ TeV. Besides, the multi charged component of the scalar multiplet, for example $\phi^{++}$ can be produced via $qq'\rightarrow W^{+} \phi^{++}\phi^{-}$ and consequently will have the cascade decay, $\phi^{++}\rightarrow \phi^{+}W^{+*}$ etc, that will lead to multi-lepton final states and missing energy. The condition $M_{\phi}>M_{F_{1}}$ then also sets $M_{\phi}\simgt 3.2$ TeV.

We scan over $M_{F_{1}}\in (1,50)$ TeV, $M_{F_{2,3}}\in M_{F_{1}}+(1,10)$ TeV, $M_{\phi}\in (10,100)$ TeV, $m_{S}\in (500\,\text{GeV},\,50\,\text{TeV})$ and $\lambda_{S}\in (0.001,0.1)$. The yukawa couplings, $f_{\alpha\beta}$ and $g_{i\alpha}$ are chosen so that they satisfy the low energy neutrino constraints. Afterwards, the rate of charged LFV processes are determined for all cases in this near degenerate limit. Although the generalized KNT model can contain a viable DM candidate, here we have studied charged LFV aspects of the model. In the companion paper \cite{talal}, we show that for standard freeze-out scenario, the DM relic density constraint leads to a very small window of mass at TeV range but if the DM content of the universe is set by non-thermal process, the constraint on the mass can be relaxed.

\subsection{Charged LFV Processes}\label{LFVprocessesresult}
\begin{figure}[h!]
\centerline{\includegraphics[width=8cm]{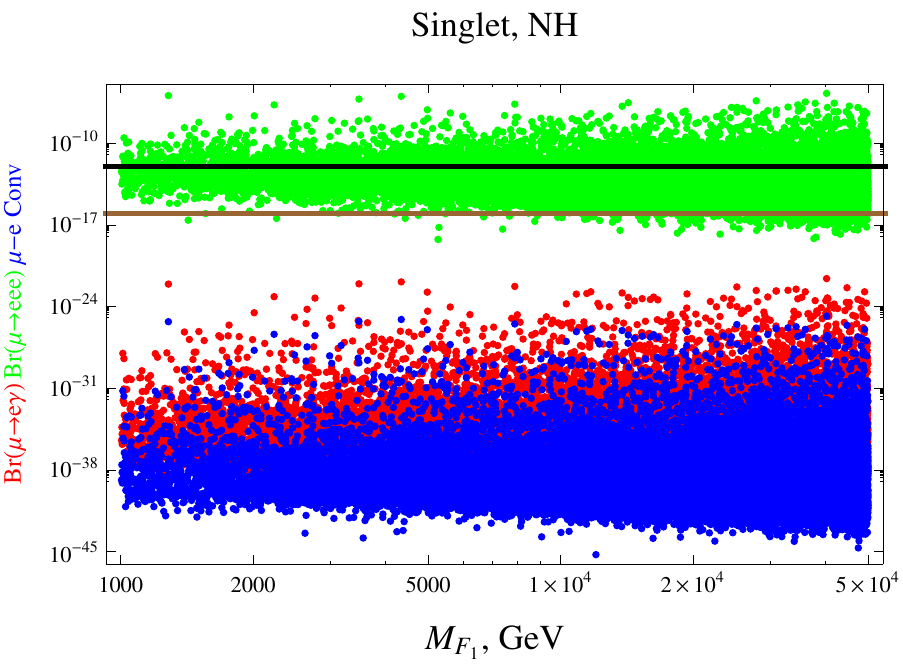}\hspace{1mm}
\includegraphics[width=8cm]{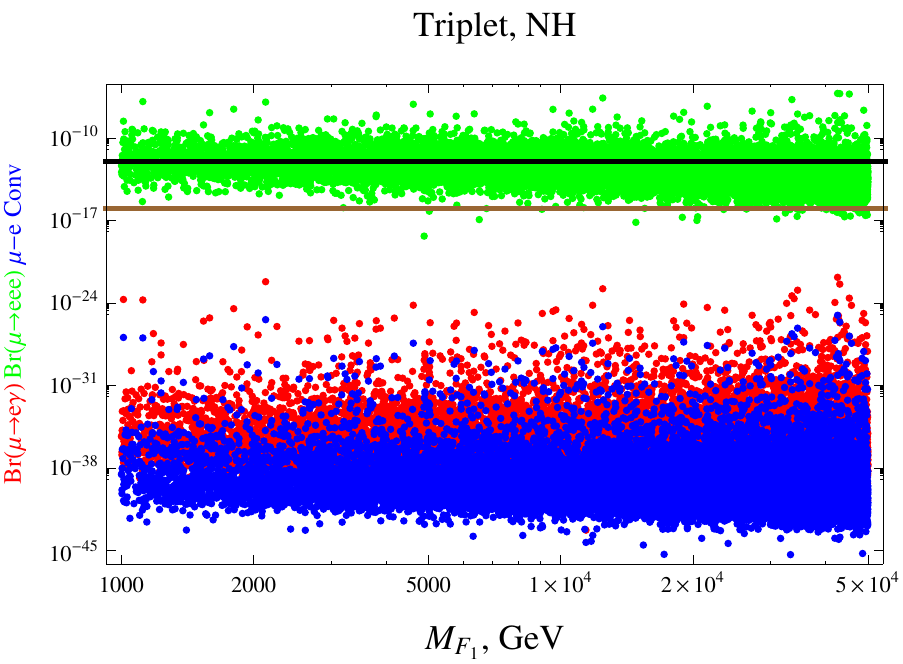}}\vspace{1mm}
\centerline{\includegraphics[width=8cm]{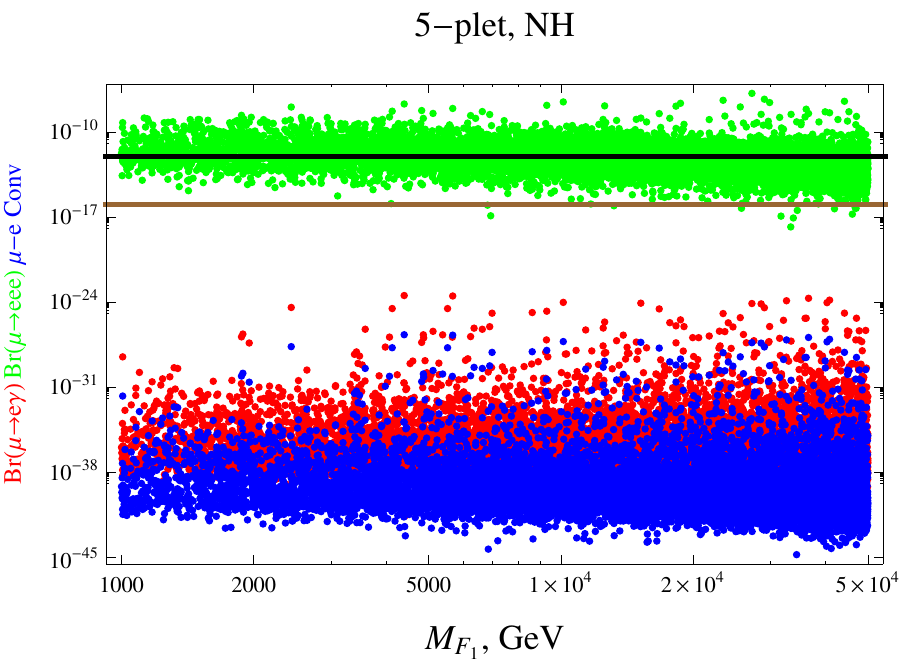}\hspace{1mm}
\includegraphics[width=8cm]{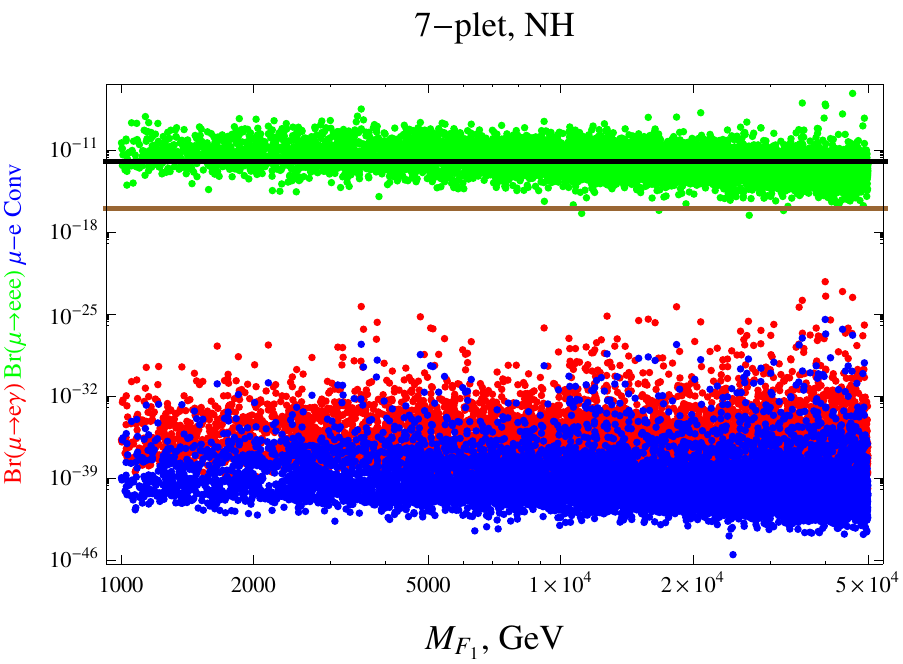}}
\caption{LFV Processes in singlet, triplet, 5-plet and 7-plet cases for normal hierarchy of neutrino masses. The graphs show same pattern for inverted hierarchy so they are not included here.}
\label{LFVNH}
\end{figure}

We can see from Fig. \ref{LFVNH} that the rate of $\mu\rightarrow e e \overline{e}$ is very large compared to the $\mu\rightarrow e\gamma$ rate and $\mu-e$ conversion rate in Au and Ti nuclei. The main reason behind this suppressed rate in $\mu\rightarrow e\gamma$ and $\mu-e$ conversion rate is the cancellations among several one-loop diagrams, as mentioned in section \ref{mutoegamma} and \ref{gammapenguin}, which have rendered dipole $A^{(2)}_{D}$ and non-dipole $A^{(2)}_{ND}$ contributions, associated with photon line attached to charged fermions, into zero in the near degenerate limit. Moreover, there are also cancellations in $A^{(1)}_{D}$ and $A^{(1)}_{ND}$ in this limit as shown in section \ref{mutoegamma} and section \ref{gammapenguin}.

\begin{figure}[h!]
\centerline{\includegraphics[width=8cm]{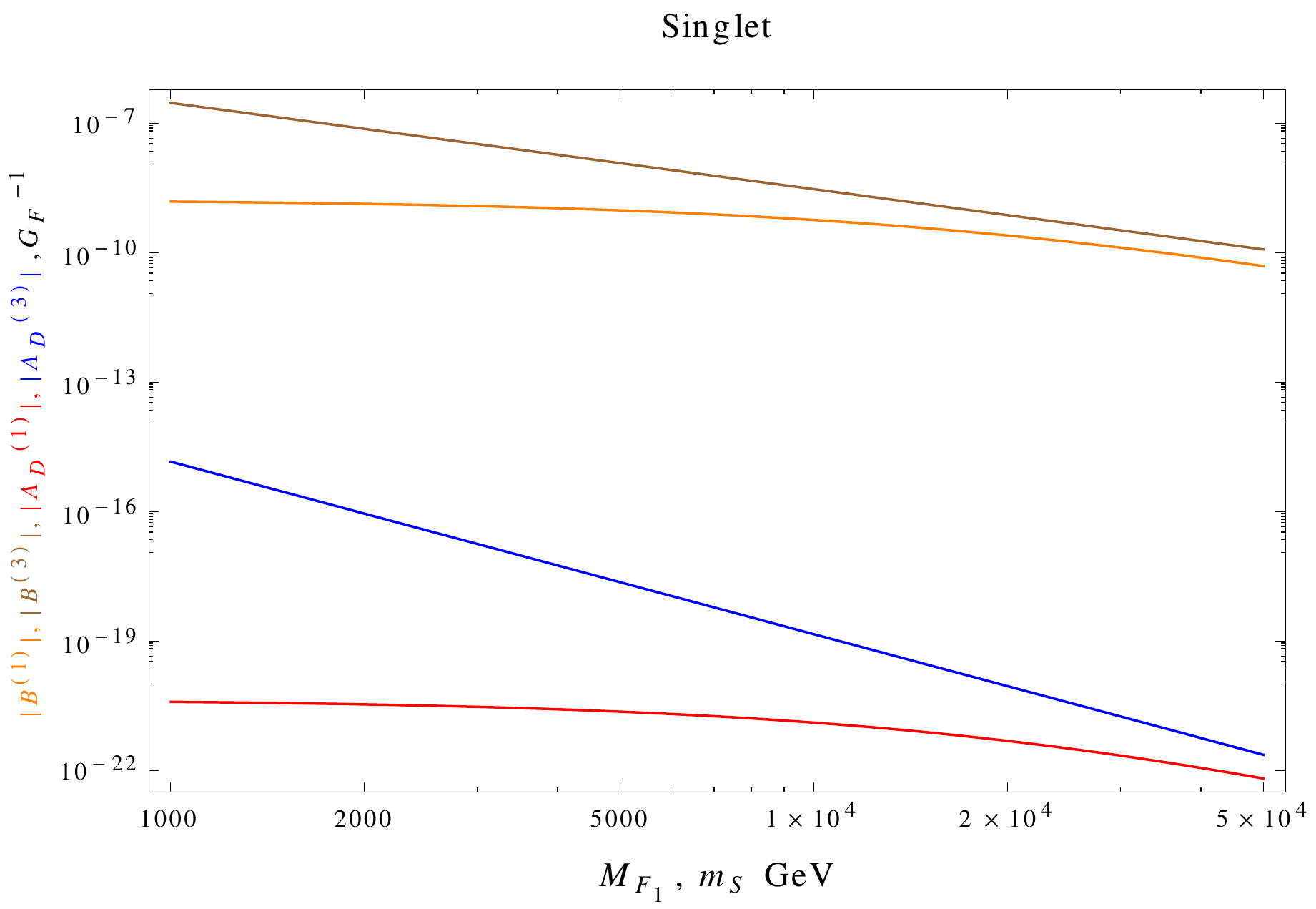}\hspace{0mm}
\includegraphics[width=8cm]{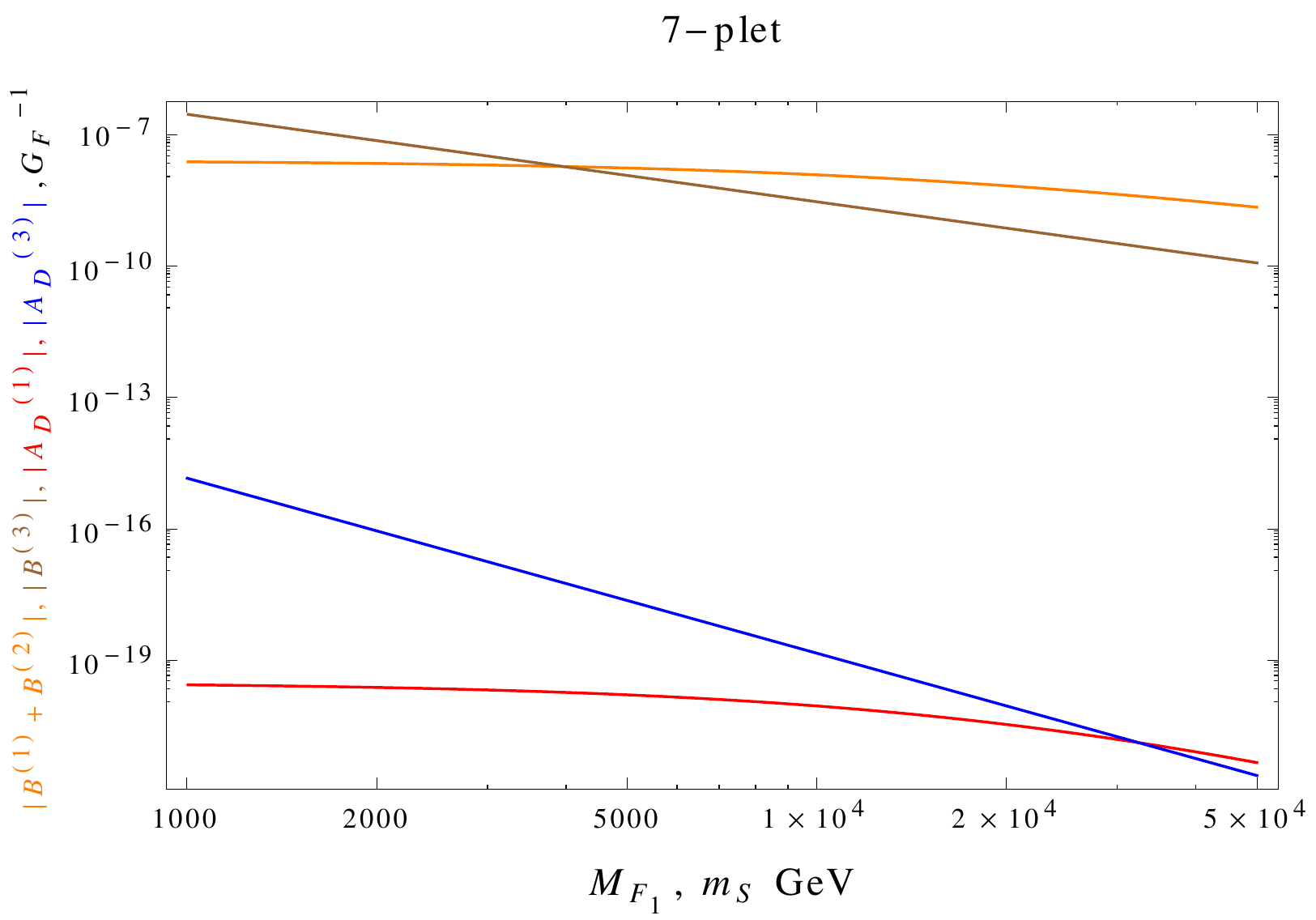}}
\caption{(Left) Relative comparison among dipole contributions, $A^{(1)}_{D}$ and $A^{(3)}_{D}$ and  box contributions, $B^{(1)}$ and $B^{(3)}$ in $G_{F}^{-1}$ unit for the singlet case. Here we can see that, box contributions are larger that dipole contributions. (Right) Similar comparison is made for the 7-plet case. As $A_{ND}$ behaves similarly as $A_{D}$ and also $F_{Z}$ is comparatively smaller than $A_{D}$ and $B$, we have not included them in the figure.}
\label{loopdig}
\end{figure}

In addition, the contribution to $Z$ penguin, $F_{Z}$ also receives several cancellations in one-loop diagrams, as mentioned in section \ref{zpenguin}. On the other hand, such cancellations does not take place in the box contribution, $B$ and in the near degenerate limit, all box diagrams coherently add up for each of the singlet, triplet, 5-plet and 7-plet cases. This can be seen from Fig. \ref{loopdig}. Consequently, $A_{D}$, $A_{ND}$ and $F_{Z}$ enter into $\mu\rightarrow e\gamma$, $\mu\rightarrow e e \overline{e}$ and $\mu-e$ conversion rates whereas $B$ also contributes into $\mu\rightarrow e e \overline{e}$ rate. Finally we can see from Fig. \ref{LFVNH} that for $M_{F_{1}}=1-50$ TeV range, part of the viable parameter space of generalized KNT model is already excluded by $\mu\rightarrow e e \overline{e}$ rate set by SINDRUM and future Mu3e experiment will exclude almost all of the parameter space for all cases within this mass range. This implies that the masses of BSM fermion and scalar particles of KNT model had to be pushed beyond $50$ TeV.

Also in Fig. \ref{loopdig} (right), for 7-plet case, the box contribution coming from diagrams with both neutral and charged fermions associated with $G$ yukawa sector, $|B^{(1)}+B^{(2)}|$ wins over $|B^{(3)}|$ associated with $F$ yukawa because, due to larger scalar and fermion multiplets, more particles enter into the loop and therefore $|B^{(1)}+B^{(2)}|$ becomes larger for $M_{F_{1}}$ than $B^{(3)}$ for $m_{S}$. Similar pattern can be seen also in dipole contributions.

\section{Conclusion and Outlook}\label{conclusion}
We have investigated charged lepton flavor violating processes $\mu\rightarrow e\gamma$, $\mu\rightarrow e e \overline{e}$ and $\mu-e$ conversion in Au and Ti in the generalized KNT model with singlet, triplet, 5-plet and 7-plet. We have shown that due to the cancellation among several one-loop contributions to photonic dipole term, photonic non-dipole term and Z-penguin term $A_{D}$, $A_{ND}$ and $F_{Z}$ respectively, the rates of $\mu\rightarrow e\gamma$ and $\mu-e$ conversion in Au and Ti become highly suppressed compared to $\mu\rightarrow e e \overline{e}$. This is due to the coherent addition of one-loop box diagrams where no cancellations take place and leads to box contribution $B$ which enters into $\mu\rightarrow e e \overline{e}$  process.
As a consequence, we have seen that for $M_{F_{1}}=1-50$ TeV mass range, the region of viable parameter space set by neutrino sector is already excluded by the limit from SINDRUM and future Mu3e will have enough sensitivity to exclude almost all of the parameter space in this mass range and thus push the mass of lightest fermionic component larger than 50 TeV in generalized KNT model.

\appendix

\section{Loop functions}\label{loopappendix}

The loop functions relevant for the dipole and non-dipole form factors from $\mu e \gamma$ vertex are
\begin{eqnarray}
 F_{1}(x)&=&\frac{1-6x+3 x^2+2 x^3-6 x^2\text{ln}x}{6(1-x)^{4}}\label{dipoleneu}\\
 F_{2}(x)&=&\frac{2+3x-6x^2+x^3+6x\text{ln}x}{6(1-x)^{4}}\label{dipolechar}\\
 G_{1}(x)&=&\frac{2-9x+18x^2-11x^3+6x^3\text{ln}x}{6(1-x)^4}\label{nondipoleneu}\\
 G_{2}(x)&=&\frac{16-45x+36x^2-7x^3+6(2-3x)\text{ln}x}{6(1-x)^4}\label{nondipolechar}
\end{eqnarray}

In the following we collect the Passarino-Veltman loop functions.
\begin{equation}
 B_{1}(m_{1},m_2)=-\frac{1}{2}-\frac{m_1^{4}-m_{2}^{4}+2m_{1}^4\text{ln}\frac{m_{2}^2}{m_{1}^{2}}}{4(m_{1}^2-m_{2}^2)^2}
 +\frac{1}{2}\text{ln}\frac{m_{2}^{2}}{\mu^{2}}
 \label{bfunction}
\end{equation}

\begin{equation}
 C_{0}(m_{1},m_{2},m_{3})=\frac{m_{2}^{2}(m_{1}^2-m_{3}^2)\text{ln}\frac{m_{2}^2}{m_{1}^2}-(m_{1}^{2}-m_{2}^{2})m_{3}^2
 \text{ln}\frac{m_{3}^{2}}{m_{1}^{2}}}{(m_{1}^2-m_{2}^{2})(m_{1}^{2}-m_{3}^{2})(m_{2}^{2}-m_{3}^{2})}
\end{equation}

\begin{align}
 C_{24}(m_{1},m_{2},m_{3})&=\frac{1}{8(m_{1}^2-m_{2}^{2})(m_{1}^{2}-m_{3}^{2})(m_{2}^{2}-m_{3}^{2})}\left[-2(m_{1}^{2}+m_{2}^{2})
 m_{3}^{4}\,\text{ln}\frac{m_{3}^{2}}{m_{1}^{2}}
 -(m_{3}^2-m_{1}^{2})\right.\notag\\
 &\left.\left(2m_{2}^{4}\,\text{ln}\frac{m_{2}^{2}}{m_{1}^{2}}
 +(m_{1}^{2}-m_{2}^{2})(m_{2}^{2}-m_{3}^{2})
 \left(2\,\text{ln}\frac{m_{1}^2}{\mu^{2}}-3\right)\right)\right]
\end{align}

\begin{eqnarray}
 \tilde{D}_{0}(m_1,m_2,m_3,m_4)&=&\frac{m_{2}^{4}\,\text{ln}\frac{m_{2}^{2}}{m_{1}^{2}}}
 {(m_{2}^{2}-m_{1}^{2})(m_{2}^{2}-m_{3}^{2})(m_{2}^{2}-m_{4}^{2})}-\frac{m_{3}^{4}\,\text{ln}\frac{m_{3}^{2}}{m_{1}^{2}}}
 {(m_{3}^{2}-m_{1}^{2})(m_{3}^{2}-m_{2}^{2})(m_{3}^{2}-m_{4}^{2})}\nonumber\\
 &-&\frac{m_{4}^{4}\,\text{ln}\frac{m_{4}^{2}}{m_{1}^{2}}}
 {(m_{4}^{2}-m_{1}^{2})(m_{4}^{2}-m_{2}^{2})(m_{4}^{2}-m_{3}^{2})}
 \label{dtilde}
\end{eqnarray}

\begin{eqnarray}
 D_{0}(m_1,m_2,m_3,m_4)&=&\frac{m_{2}^{2}\,\text{ln}\frac{m_{2}^{2}}{m_{1}^{2}}}
 {(m_{2}^{2}-m_{1}^{2})(m_{2}^{2}-m_{3}^{2})(m_{2}^{2}-m_{4}^{2})}-\frac{m_{3}^{2}\,\text{ln}\frac{m_{3}^{2}}{m_{1}^{2}}}
 {(m_{3}^{2}-m_{1}^{2})(m_{3}^{2}-m_{2}^{2})(m_{3}^{2}-m_{4}^{2})}\nonumber\\
 &-&\frac{m_{4}^{2}\,\text{ln}\frac{m_{4}^{2}}{m_{1}^{2}}}
 {(m_{4}^{2}-m_{1}^{2})(m_{4}^{2}-m_{2}^{2})(m_{4}^{2}-m_{3}^{2})}
 \label{dzero}
\end{eqnarray}

\end{document}